\documentclass[aps,twocolumn,superscriptaddress,floatfix]{revtex4}
\usepackage{graphicx,psfrag,amsmath,amssymb,amsfonts,latexsym,color,dcolumn,bm}

\begin{document}
	
\title{Scaling laws for harmonically trapped two-species mixtures at thermal equilibrium}

\author{Francisco Jauffred}

\affiliation{\mbox{Department of Physics, University of Massachusetts, Boston, MA 02125, USA}}

\author{Roberto Onofrio}

\affiliation{\mbox{Dipartimento di Fisica e Astronomia ``Galileo Galilei'', Universit\`a  di Padova, 
		Via Marzolo 8, Padova 35131, Italy}}

\affiliation{\mbox{Department of Physics and Astronomy, Dartmouth College, 6127 Wilder Laboratory, 
		Hanover, NH 03755, USA}}

\author{Bala Sundaram}

\affiliation{\mbox{Department of Physics, University of Massachusetts, Boston, MA 02125, USA}}

\begin{abstract}
We discuss the scaling of the interaction energy with  particle numbers for a harmonically trapped
two-species mixture at thermal equilibrium experiencing interactions of arbitrary strength and range.
In the limit of long-range interactions and weak coupling, we  recover known results for
the integrable Caldeira-Leggett model in the classical limit.
In the case of short-range interactions and for a balanced mixture, numerical simulations
show scaling laws with exponents that depend on the interaction strength, its attractive
or repulsive nature, and the dimensionality of the system.
Simple analytic considerations based on equilibrium statistical mechanics and
small interspecies coupling quantitatively recover the numerical results. 
The dependence of the scaling on interaction strength helps to identify a threshold between two distinct regimes.
Our thermalization model covers both local and extended interactions allowing for interpolation between different
systems such as fully ionized gases and neutral atoms, as well as parameters describing integrable and chaotic dynamics.
\end{abstract}
\date{\today}
\maketitle

\section{Introduction}

Thermalization in many-body systems is a topic of broad interest in a variety of
contexts including fluids, plasmas, and chemical reaction dynamics.
An approach which has been considered of universal character, as it may be used for both classical and quantum
systems, is based on a closed Hamiltonian dynamics and referred to as the Caldeira-Leggett model, though its
origin can be traced to earlier contributions \cite{Magalinskii,Ullersma,Caldeira1,Caldeira2,Caldeira3}.

In previous work~\cite{OnoSun,JauOnoSun1}, we explored thermalization in the context of a model where the
interaction, both in range and strength, appeared as a generalization of this more familiar Caldeira-Leggett 
model. Though the earlier context was the sympathetic cooling of atomic gas mixtures, our model was also 
intended to explore the realm of nonlinearities arising in either, or both, interaction and confining 
potentials~\cite{JauOnoSun2}. In particular, plasma physics offers a phenomenological platform to discuss 
our model as scaling properties, turbulence, strong coupling, and exothermic reactions all play a crucial role.

An intriguing feature reported in Ref.~\cite{JauOnoSun1} was power-law scaling of the average total interaction
energy with total number of particles, for equal number mixtures, as thermalization was approached.
Specifically, the scaling exponent was reminiscent of that associated with Kolmogorov scaling associated with
turbulent mixing in fluids. Within the explored range of parameters, the scaling was persistent with changing
dimensionality of the dynamics. The suggested analogy between turbulent mixing and thermalization originates
from the common issue of homogenization. As we show, the interaction energy between the two species transitions
from a dynamical regime to one more attuned to a statistical analysis. Thermalization of the two species
coincides with the realization of this latter regime, and we interpret this to be the well-mixed state. 

In this paper, we explore in more detail this scaling behavior. Aside from exploring a wider range
of parameters in our numerical simulations, we construct analytic estimates of the scaling exponents 
from various thermodynamic perspectives and with variable dimensionality. 
In one-dimension, there do exist conditions under which the exponent does indeed coincide 
with that seen in Kolmogorov scaling while, under analogous conditions, there are deviations at higher 
dimensionality. It is worth noting that dimensional arguments behind Kolmogorov scaling are scalar
in nature, due to assumptions of isotropy, resulting in an effective one-dimensionality.
In our dynamical situation, one-dimensionality favors energy transfer due to the absence of constraints
on head-on collisions while, in higher dimensions, angular momentum serves to restrict these, resulting
in a slower rate of energy transfer. Also, the scaling becomes extensive in the limit of infinite dimensions,
as expected from mean-field constructs and phase space considerations.

On exploring a broader range of parameters, scaling indicates a saturation in the total interaction energy.
This saturation phenomenon occurs when the interaction range is so large that all possible pairwise interactions occur,
regardless of the interaction strength. Analogously, saturation also occurs for any interaction range when the interaction
strength is so large that either clustering of particles of different species (when their interaction is attractive),
or nearly complete spatial separation (when the interaction is repulsive) occurs. The phenomenon is microscopically
related to the two-point spatial correlation function between particles of different species.
This effect also implies that in a more general setting in which both attractive and repulsive interactions
occurs, such as in dense plasmas in the strong coupling regime, the attractive component dominates over the repulsive
component in establishing the dynamics and the total interaction energy.
Further relationships of our work to plasma physics, with particular regard to possible future research
directions on anisotropic turbulence and efficient heating protocols, are highlighted in the conclusions.

\begin{figure*}[t]
\includegraphics[width=0.95\columnwidth, clip=true]{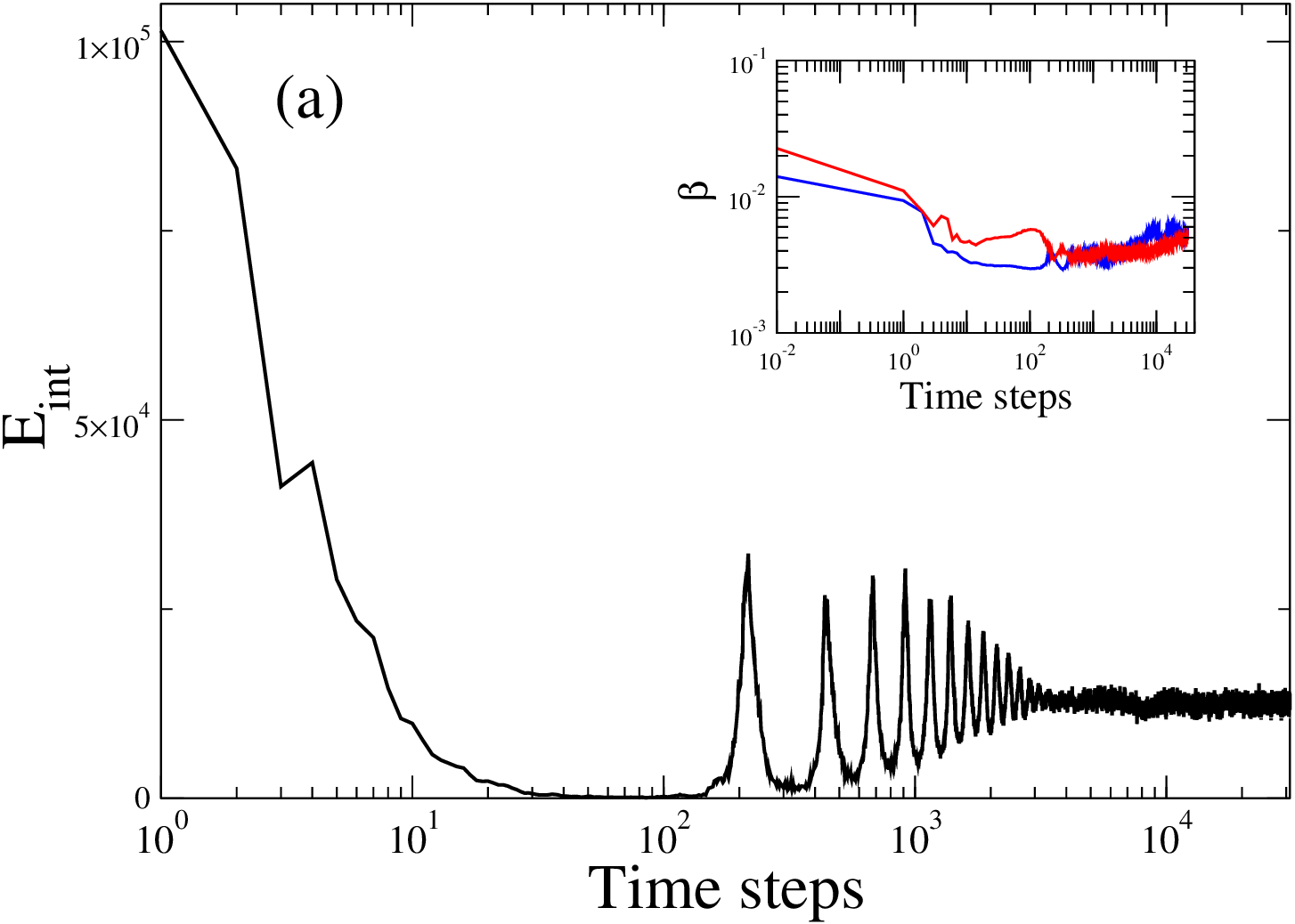}
\includegraphics[width=0.95\columnwidth, clip=true]{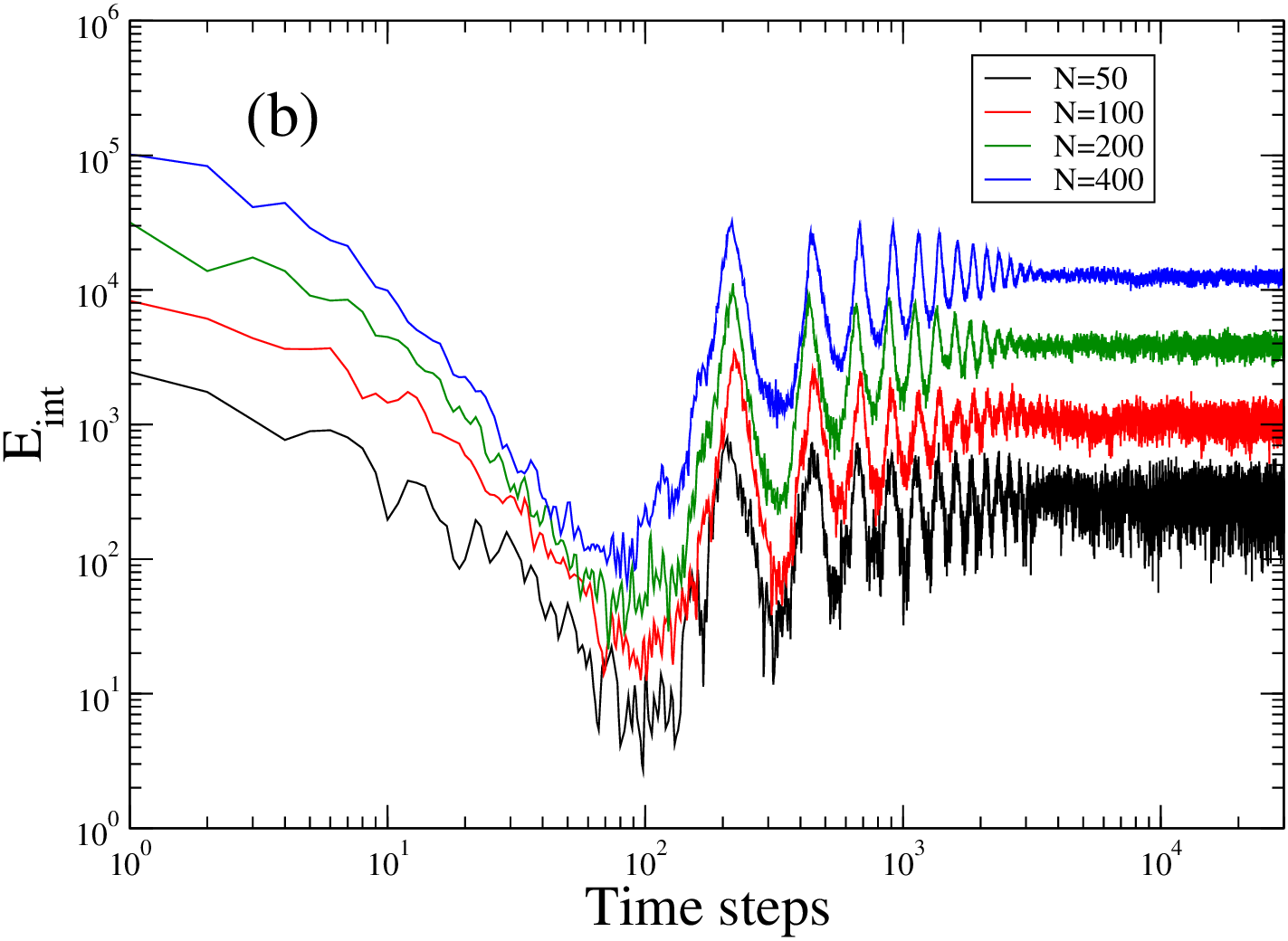}
\caption{Left: Plot of the total interaction energy versus time for a given equal number of particles in the two systems. 
In the inset the inverse temperatures of the two systems are shown versus time, confirming the presence of a regime
in which thermalization is assured, and showing also the presence of an exothermic equilibration.
The simulations were performed with $N_A=N_B=400$ particles, $\gamma$=20.0, $\lambda$=0.1, $\beta_A$=2.0, $\beta_B=0.2$,
$m_A=m_B=1.0$, $\omega_A=1.0$, $\omega_B=144/89$. Right: Same quantity for varying number of particles in the two systems,
qualitatively showing that the interaction energy increases with the number of particles, and that its fluctuations
after the thermalization stage decrease with the number of particles.}
\label{Fig1}
\end{figure*}

\section{A generalized interaction model between two species}

The Hamiltonian we consider is~\cite{OnoSun,JauOnoSun1,JauOnoSun2}
\begin{eqnarray}
H&=&\sum_{m=1}^{N_A} \left(\frac{P_m^2}{2m_A}+\frac{1}{2} m_A \omega_A^2 Q_m^2 \right)+ \nonumber \\
& & \sum_{n=1}^{N_B} \left(\frac{p_n^2}{2m_B}+\frac{1}{2} m_B \omega_B^2 q_n^2 \right)+  \nonumber \\
& & \gamma \sum_{m=1}^{N_A} \sum_{n=1}^{N_B} \exp{\left[-\frac{\left(Q_m-q_n\right)^2}{\lambda^2}\right]},
\label{Hamilton}
\end{eqnarray}
where $(Q_m,P_m)$ and $(q_n,p_n)$ are the positions and momenta of each particle of the
two species $A$ and $B$, respectively, and positions lie in a generic $D$-dimensional space.
The interspecies term is governed by two parameters, the strength $\gamma$ and the
range $\lambda$ of the interaction, with the former representing the typical energy exchanged between
two distinct particles in a close-distance interaction. 
Although the interaction Hamiltonian looks rather simple, it allows for the study of a variety of situations, including
balanced ($N_A=N_B$) and unbalanced mixtures, attractive ($\gamma<0$) and repulsive ($\gamma>0$) interactions, as well
as long-range ($\lambda \rightarrow \infty$) and short-range interactions. 
In particular, for a completely unbalanced mixture (for instance $N_A=1$ and $N_B \rightarrow \infty$),
small $\gamma$ and large interaction range, the model mimics, in the classical limit and for a finite number of
particles~\cite{Smith}, the genuine Caldeira-Leggett approach used to model dissipation in open systems. 
It should be noted that we are considering the classical dynamics so reference to Caldeira-Leggett 
is in terms of the  functional form of the interaction.
Our choice of a Gaussian form of the interaction term in the Hamiltonian allows for simple analytical
estimates based on the canonical ensemble, and therefore in the thermodynamic limit of infinite particle numbers. As first
emphasized in~\cite{Khilchin}, thermalization and equilibration processes are quite insensitive to the microscopic details of the
interaction. Therefore we expect our results to be relevant beyond the specific, analytically convenient, interaction
term we have adopted.

The equations of motion corresponding to the Hamiltonian in Eq.~\ref{Hamilton} can be numerically integrated 
to machine precision. Given the harmonic trapping potential, the initial conditions are drawn from canonical 
energy distributions consistent with the initial temperatures of the two clouds (see ~\cite{JauOnoSun1} for details). 
The time evolution of the particle trajectories, in the presence of interactions, allows us to track the dependence 
of the total interaction energy on time for typical parameters as shown in Fig.~\ref{Fig1}. 
Early on, the interaction energy reflects the periodicity associated with the harmonic trap, while increasing 
aperiodicity develops with time. For even longer times, the interaction energy settles into a noisy time-averaged 
value. The inset shows the evolution of the inverse temperatures $\beta_A$ and $\beta_B$ of the two subsystems over the
same time, with equilibration coinciding with the settling down of the interaction energy.
The inverse temperature, as discussed in detail in Ref.~\cite{JauOnoSun1}, is evaluated by looking at the
energy variance $\sigma_E^2$ where

\begin{equation}
\sigma_E=(\langle E^2 \rangle-\langle E \rangle^2)^{1/2}=\sqrt{D}/\beta,
\label{inversebeta}
\end{equation}
where $D$ is the spatial dimensionality, and the averages are taken over the ensemble of particles at any given time.
This simple relationship relies on Gibbs-Boltzmann statistics, and therefore a weak-coupling approximation
~\cite{Montroll,AndersenK,AndersenH}. The interaction energy is a more robust, coarse-grained
indicator whose validity also holds in the strong coupling limit. More specifically, we have observed situations,
for instance due to strong interspecies repulsion, for which the system does not settle into an equilibrium state
as defined by a common inverse temperature, yet a stationary situation occurs with different effective inverse
temperatures as evaluated from Eq.~\ref{inversebeta} and a stationary interaction energy. 
By repeating the numerical simulations for different numbers of particles, we can highlight the scaling behavior
seen in the late time-averaged, total interaction energy with respect to the number of particles, reported in Ref.~\cite{JauOnoSun1}.
Note that for the rest of the paper, the term {\it total interaction energy}, denoted by 
$\bar{E}_{\mathrm{int}}$, will refer to the post saturation, time-averaged total interaction energy. 
In the scaling context, the right panel of Fig.~\ref{Fig1} notes both the change in this average 
energy as well as the reduction in fluctuations with increasing particle number. 
This will prove relevant in determining the accuracy of the power-law scaling results with changing system size.

\begin{figure}[t]
\includegraphics[width=0.95\columnwidth, clip=true]{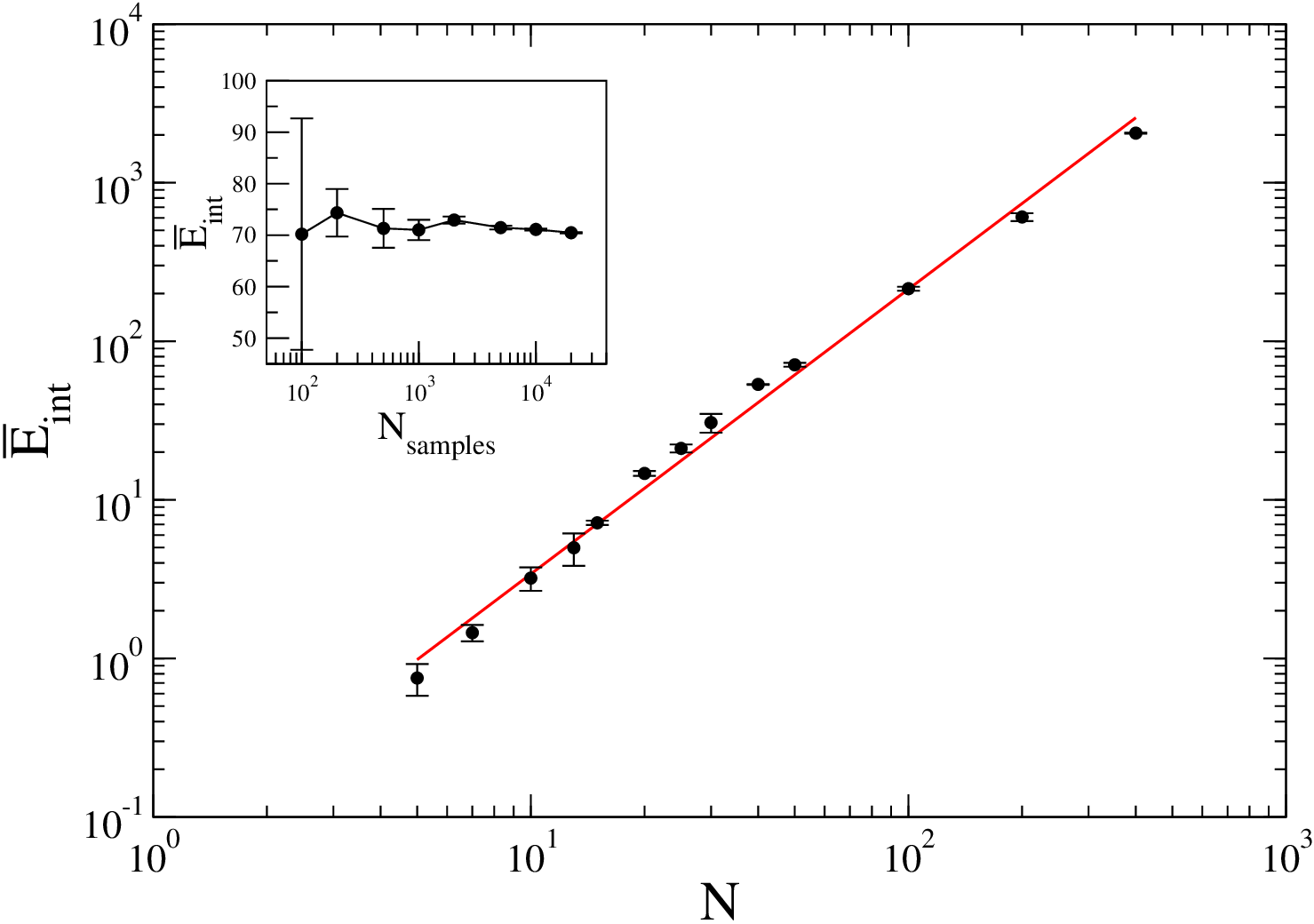}
\caption{Total interaction energy $\bar{E}_{\mathrm{int}}$ versus the number of particles $N$, with both quantities 
shown on logarithmic scales, for a broad range of $N$ encompassing also the few-body case.
Deviations from a power-law scaling at small number of particles are attributable 
to the few-body dynamics which does not allow for meaningful comparison of the time-averaged total interaction 
energy to the corresponding one obtained by a canonical ensemble average.
Including points from the few-body cases does indeed affect the best fit significantly.
A scaling exponent $\alpha=1.80 \pm 0.04$ is obtained for a global fit (red line).
Considering only the four rightmost points gives $\alpha=1.61 \pm 0.04$, while the eight rightmost
result in $\alpha=1.62 \pm 0.02$, showing robustness of the fit for large $N$.
The inset shows the dependence of the standard deviation of the total interaction energy on the number of
time steps used for evaluating its average value at the end of the simulation, for the
case of $N$=50 particles. The optimal choice is a compromise
between the larger standard deviations for smaller time sequences, and the need to avoid bias due
to possible residual thermalization dynamics for larger size of the sample.
The coupling strength is $\gamma=2.0$, while the parameters $\lambda, \beta_A, \beta_B, m_A, m_B, \omega_A, 
\omega_B$ have the same values as in Fig.~\ref{Fig1}.
The error bars in the inset correspond to one standard deviation from the average value, 
while the errors on the scaling exponent $\alpha$ here and in the following figures 
are evaluated as one standard deviation in the least squares analysis.
Based on this analysis, we use a minimum of $N$=50 particles for each
species, and $10^3$ time steps for the time averaging in all other figures.}
\label{Fig2}
\end{figure}

\section{Numerical Exploration}

We have numerically evaluated the critical exponents for a range of parameters, and for balanced systems with
$N=50, 100, 200, 400, 800$ particles each. The acceptable lower number of particles is determined by the large 
statistical fluctuations of the interaction energy, while the higher number is limited by the duration of the
simulation (requiring about two weeks for the largest number of particles, $N=800$, and $10^5$ time steps on
a single processor). Although, in many cases, thermalization can occur on shorter timescales, we have decided to 
standardize the simulations by considering a total duration of $10^5$ time steps in all cases.
The interaction energy is evaluated by averaging over the last $10^3$ time steps of each simulation, where
the discussion of the inset plot in the caption of Fig.~\ref{Fig2} provides justification for this choice.

\begin{figure*}[t]
\includegraphics[width=0.95\columnwidth, clip=true]{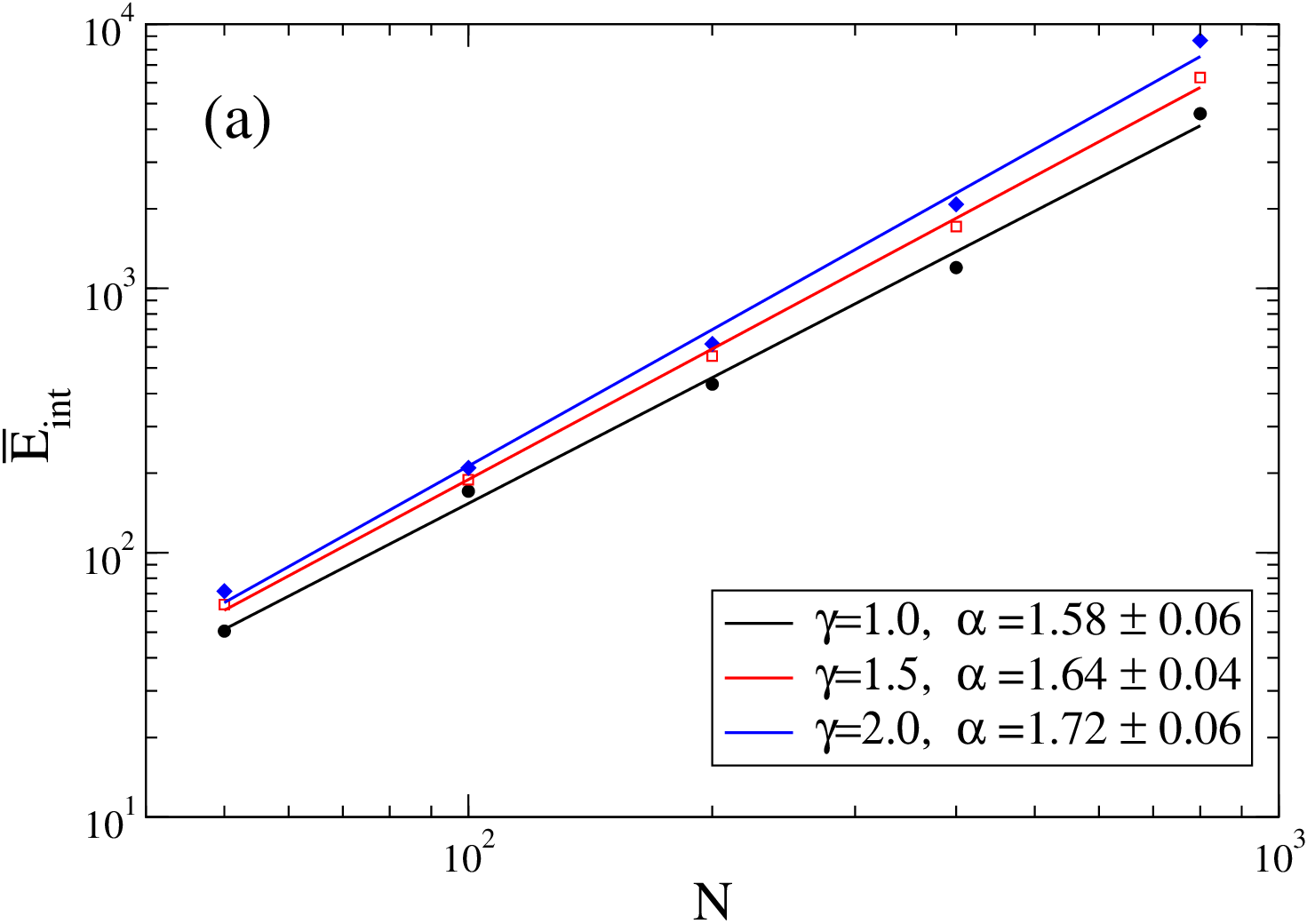}
\includegraphics[width=0.95\columnwidth, clip=true]{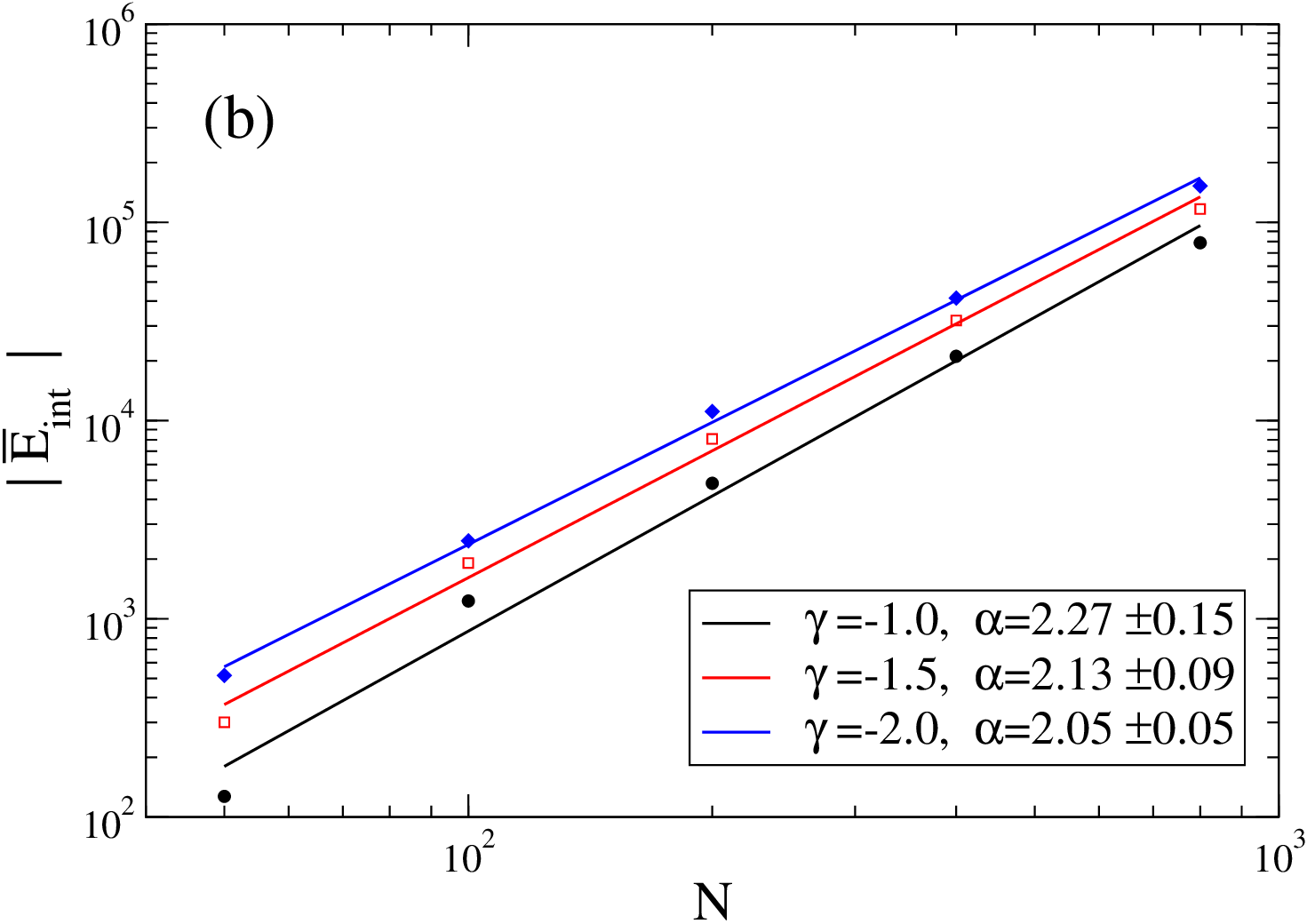}
\caption{Scaling of the total interaction energy with system size ($N_A=N_B=N$) for both
repulsive (left) and attractive (right) cases. The strength $\gamma$ is specified while
$m_A=m_B=1.0$, $\lambda=0.1$, $\omega_A= 1.0,\omega_B=144/89$, $\beta_A=0.2$ and $\beta_B=2.0$.}
\label{Fig3}
\end{figure*}

In our attempts to improve the precision of the power-law exponent, we have extensively
studied its dependence on the run time, the series length of the time-dependent interaction energy
used in the time average, as well as the number of particles.
The latter is a crucial parameter because we expect that for small number
of particles the ergodic hypothesis does not hold on the limited timescales we explore.
A manifestation of this can be seen in Fig.~\ref{Fig2} where the variation of the time
averaged interaction energy with particle number is shown.
Deviations from the power law behavior seen for the smaller $N$ values are a consequence
of the absence of ergodicity on the timescales of our simulations.
By allowing enough time for thermalization and optimizing the  averaging time windows in the thermalization regime,
as indicated in the caption for Fig.~\ref{Fig2}, we obtain the accuracy necessary (relative errors of a few percentages) to 
validate the predicted behavior for the scaling exponent. 

In Fig.~\ref{Fig3} we show the scaling of this time-averaged interaction energy, in one-dimension, with the number
of particles for various $\gamma$, corresponding to repulsive and attractive interactions.
The values of $\gamma$ lie between the perturbative case (where thermalization occurs 
on exceedingly long time scales but there is some analytic tractability) and the strongly coupled case, 
where any analytical perturbative construction is not expected to hold. On comparing the repulsive and 
attractive cases in Fig.~\ref{Fig3}, if all the other parameters are kept equal, then the interaction energy 
(absolute value) for the attractive case is at least one order of magnitude larger with respect to the repulsive case. 
This nontrivial feature may be interpreted as due to the different role 
played by the interaction term in the two cases. The textbook scenario of thermalization consists of two compartments of 
particles where the interaction (wall between them) is very weak, either due to the interparticle interactions themselves 
or because the interface between the subsystems is of lower dimensionality as compared with those of the sub-systems. 
The repulsive case follows this scenario. However, this is not a likely scenario for attractive interactions where aggregation or 
clustering can lead to increased strength in the spatially dependent interactions. Using this notion, the total interaction energy 
is bounded simply by $E_{\mathrm{int}}^{\mathrm{sat}}=-\gamma N_A N_B$, when the distance between all pairs $|Q_m-q_n| << \lambda$. 
Including the fact that the particles are moving means that the asymptotic interaction energy seen in the numerics is considerably 
less (in absolute value). The reduction factor can be estimated, in the case of small $\lambda$, by comparing the timescale on which 
the two particles are proximal, i.e. within $\lambda$ (of order $\lambda/v$ where $v$ is their relative velocity) with the period of 
the harmonic oscillation in the trap. For the parameters in Fig.~\ref{Fig3}, the typical saturation value is about 10
$\%$ of $E_{\mathrm{int}}^{\mathrm{sat}}$. 

Also, the attractive case is more efficient, for the same choice of initial conditions, in increasing the total interaction energy of the two systems. 
The interface between the two systems is more extended in configurational space and there is aggregation rather then phase separation 
as in the repulsive case. As a consequence, the interactions proceed faster and involve larger clusters of particles. 
Conversely, as shown by numerical simulations and simple analytical estimates, thermalization in the case of strong repulsion 
occurs intermittently as it involves small particle numbers at the tails of the already phase-separated clouds. 
In the attractive case, the thermalization phenomenon can be viewed as proceeding through latent energy stored in the interaction term, which 
is then released as kinetic and potential energy of each particle. In this sense, it can be viewed as a generalization of Joule-Thompson 
effects in real gases, with a compression and heating stage rather than the usual expansion and cooling.

\begin{figure}[t]
\includegraphics[width=0.95\columnwidth, clip=true]{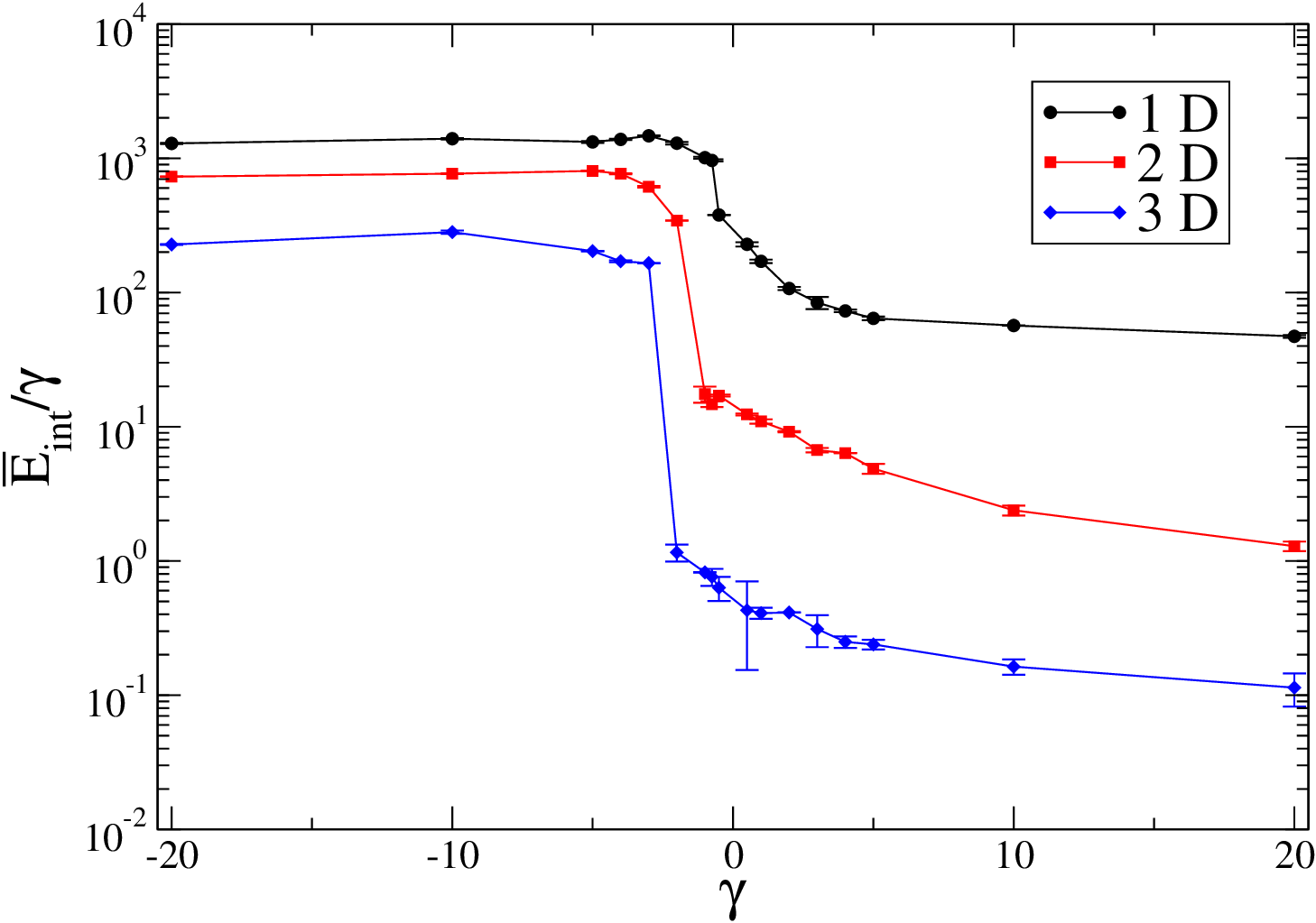}
\caption{Total interaction energy at thermal equilibrium per unit of coupling strength $\gamma$ versus the coupling strength itself for
baths made of $10^2$ particles each, and the same temperatures and interaction range as in Fig.\ref{Fig3}. 
The plots show evidence of the saturation of the interaction energy in both extremes of strong attractive and repulsive couplings in all dimensions.}
\label{Fig4}
\end{figure}

The view suggested above is further corroborated by inspecting the total interaction energy at equilibrium, normalized to the coupling strength 
$\gamma$, as a function of $\gamma$, shown in Fig.~\ref{Fig4}. The interaction energy saturates both at large values of $\gamma$ due to 
species separation for repulsive interactions, at a small value for $E_{\mathrm{int}}$, and at large negative values of $\gamma$ due to species 
clustering, with a large absolute value of $E_{\mathrm{int}}$. As discussed in~\cite{JauOnoSun1}, the interaction energy is a macroscopic
indicator of the ensemble-averaged distance between two different species particles, as
\begin{equation}
\langle (q_n-Q_m)^2 \rangle = - \lambda^2 \ln \left( \frac{\langle E_{\mathrm{int}} \rangle}{\gamma N_A N_B} \right).
\end{equation}  

In Fig.~\ref{Fig4} there is an intermediate region of values of $\gamma$ where these appears to be a crossover between
the two extreme values. As we will describe, this is where the considerations of the analytical model we develop may apply.
It would appear that special care is required in the limit as $\gamma$ approaches zero, as the interaction energy is zero
by definition in the limit. The numerical analysis has been repeated in higher dimensions, confirming
the general trend with some distinguishing features. For the same parameters, higher dimensions show ever smaller
interaction energy, as the particles may dilute in a progressively larger phase space. Also, the presence of angular momentum
allows for evasive trajectories which are forbidden in the one dimensional case. The difference between attractive and
repulsive interactions is amplified by higher dimensionality, and in the full three-dimensional (3D) case the asymptotic values of the
interaction energies differ by about three orders of magnitude, in contrast to a single order of magnitude for the 1D case. 

\begin{figure}[t]
\includegraphics[width=0.95\columnwidth, clip=true]{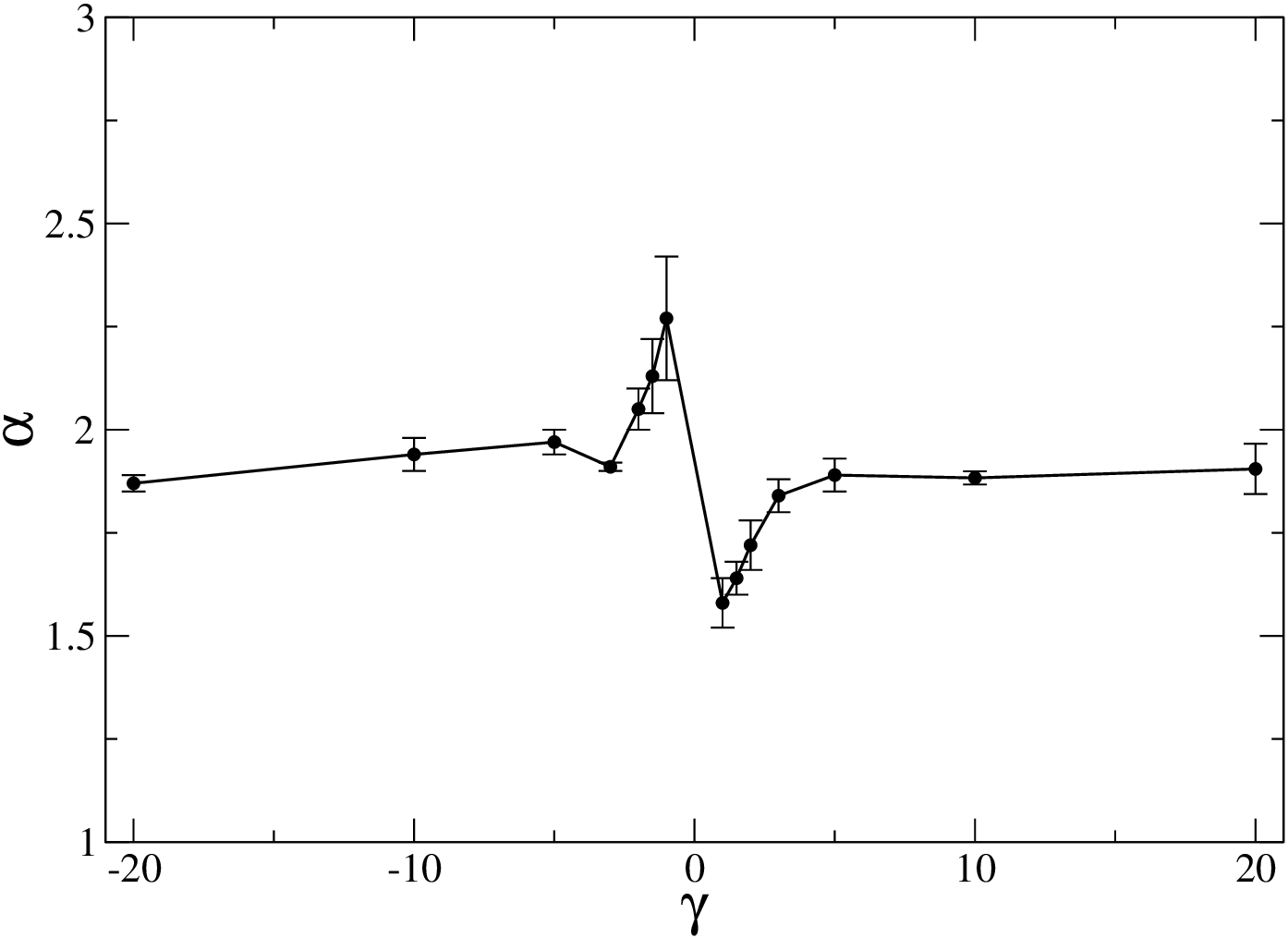}
\caption{Scaling exponent $\alpha$ versus the coupling strength $\gamma$ for the same temperatures and interaction
range as in Fig. 4, in the 1D case. A narrow region at small values of $\gamma$ is visible in which anomalous scaling occurs.
Notice that the error bars in the region of small and negative $\gamma$ are large enough to make the values compatible with
$\alpha=2$ within three standard deviations at most, while the case of anomalous scaling is statistically much stronger for
positive values of $\gamma$. Specifically, for our data, at $\gamma=1$ we get $\alpha=1.58 \pm 0.06$ which is about $1.3$ standard
deviations from the theoretically expected value $5/3$ discussed in the analytical section.
The analysis has been repeated for the case of 2D and 3D systems at different coupling strenghts and all the other parameters
kept constant as in the 1D case, obtaining exponents of $\alpha=1.85 \pm 0.02 ~(\gamma=1)$, $\alpha=1.74 \pm 0.05 ~(\gamma=2)$,
$\alpha=1.82 \pm 0.04 ~(\gamma=20)$, for the 2D case, and $\alpha=2.03 \pm 0.04 ~(\gamma=1)$, $\alpha=2.03 \pm 0.04 ~(\gamma=2)$,
$\alpha=2.26 \pm 0.12 ~(\gamma=20)$, for the 3D case. A comprehensive analysis of anomalous scaling for the higher-dimensionality
cases will be the subject of future investigation, including the case of anisotropic trapping.}
\label{Fig5}
\end{figure}

A second prominent feature in comparing the attractive and repulsive cases in Fig.~\ref{Fig3} is that the scaling
exponent is compatible with $\alpha=2$ within two standard deviations for the attractive case, and instead assumes
values significantly lower in the repulsive case. 
This suggests the consideration of a broader range of $\gamma$ values (as in Fig.~\ref{Fig4}).
The resulting dependence of the scaling exponent on coupling strength $\gamma$ is shown in Fig.~\ref{Fig5}. 
At large absolute values of $\gamma$ the scaling  exponent is compatible with $2$. In this highly nonperturbative regime, as 
noted above, the particles are strongly clustered in the attractive case, and they all interact with each other.
In the repulsive case there is species separation so we expect only intermittent interactions by particles at the boundary 
between the two separated species. This constitutes a small subset of each species, and as discussed above the 
interaction energy should therefore scale with the square of the particle number (for a balanced mixture) times a suppression 
factor proportional to the thickness of the boundary region with respect to the interaction range $\lambda$.
In the weakly interacting regime, the scaling exponent is in line with the expectations of homogeneity and
Kolmogorov-like mixing as discussed in the next section. By contrast, at large $\gamma$, the strong interparticle
interaction is analogous to a high viscosity regime in fluids, which precludes turbulence and the associated scaling.
Once again, the trend is confirmed  in higher-dimensional cases, as indicated by the data discussed in the caption of Fig.~\ref{Fig5}.

\section{Analytical Considerations}

It turns out that much of the behavior seen can be recovered using equilibrium statistical mechanics and
thermodynamics considerations. We begin by rewriting the Hamiltonian Eq.~(\ref{Hamilton}) as the sum of 
the free and interaction Hamiltonians, respectively,  $H=H_0+H_{\mathrm{int}}$. 
Having in mind weak-coupling, perturbative expansions, we make explicit the interaction strength $\gamma$ 
in the interaction Hamiltonian, such that $H_{\mathrm{int}}=\gamma I$, where $I$ is a dimensionless quantity.
The corresponding partition function and the expectation value of energy at thermal equilibrium corresponding to 
inverse temperature $\beta$ are respectively

\begin{equation}
Z=\int \prod_{m=1}^{N_A} d\vec{Q}_{m} d\vec{P}_{m}
\prod_{n=1}^{N_B} d\vec{q}_{n} d\vec{p}_{n}
\exp{\left[-\beta(H_0+\gamma I)\right]},
\label{Partition}
\end{equation} 

\begin{eqnarray} 
\langle E \rangle &=&-\frac{1}{Z} \frac{\partial Z}{\partial \beta} =  
\frac{1}{Z} \int \prod_{m=1}^{N_A} d\vec{Q}_{m} d\vec{P}_{m}
\prod_{n=1}^{N_B} d\vec{q}_{n} d\vec{p}_{n} \times \nonumber \\
& & \left(H_0 + \gamma I\right) \exp\left[-\beta (H_0+\gamma I)\right]. 
\label{PartitionEnergy} 
\end{eqnarray} 

We expand the expression  for the energy in terms of the coupling strength $\gamma$, to obtain

\begin{equation} 
\left\langle E\right\rangle =
\frac{D\left(N_A +N_B \right)}{\beta}+\gamma F_D(\rho_A,\rho_B)N_A N_B,
\label{EnergyApprox} 
\end{equation}
where we have introduced a form factor $F_D(\rho_A,\rho_B)$, defined as

\begin{eqnarray}
& & F_D(\rho_A,\rho_B)=
{\left(1+\rho_A^2+\rho_B^2\right)}^{-\frac{D}{2}} \times \nonumber \\ 
&& \left[D+1-\frac{D}{2} \frac{\rho_A^{-2} (1+\rho_A^{-2})+\rho_B^{-2} (1+\rho_B^{-2})}{(1+\rho_A^{-2})(1+\rho_B^{-2})-1}\right].
\label{FormFactor} 
\end{eqnarray} 
Here $\rho_A=\xi_A/\lambda$, $\rho_B=\xi_B/\lambda$, where $\xi_A=\sqrt{2/\beta m_A \omega_A^2}$,
and $\xi_B=\sqrt{2/\beta m_B \omega_B^2}$ are the thermal lengths of the two species. 
The analysis can be simplified by assuming that both species have equal mass and, hence, identical frequencies
in the harmonic trap, which implies $\xi_A =\xi_B=\xi$ corresponding to the inverse equilibrium temperature $\beta$.
The form factor can also be re-expressed contrasting $\beta$ with $\beta_{\lambda}$ defined as
 $\beta_\lambda =2/(m\omega^2 \lambda^2)$, which corresponds to $\xi=\lambda$.

\begin{figure}[t]
\includegraphics[width=0.95\columnwidth, clip=true]{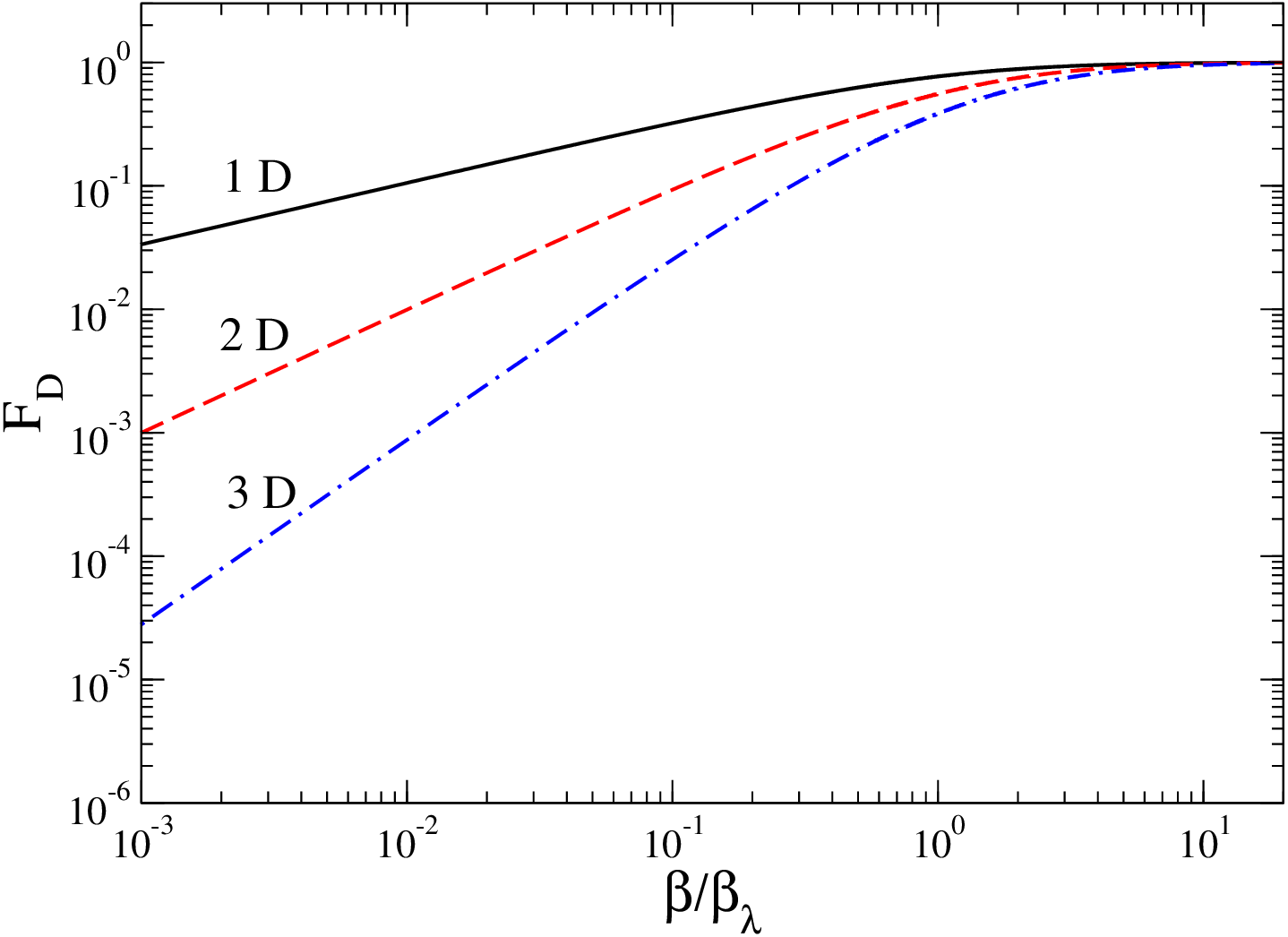}
\caption{Form factor $F_D$ versus the inverse temperature $\beta$ normalized to
the interaction range inverse temperature $\beta_\lambda$. }
\label{Fig6}
\end{figure}

Fig.~\ref{Fig6} shows the variation of the form factor with changing $\beta$ normalized to $\beta_\lambda$.
We can now differentiate behavior according to the importance of these thermal lengths with respect to the
interaction range, obtaining approximate analytical expressions for the different regimes visible in Fig. \ref{Fig6}.
This is facilitated by considering the expression for $F_D$ when $\rho_A=\rho_B=\rho$:
\begin{equation}
F_D(\rho)=\frac{ \rho^{-2} (1+2\rho^2)^{-\frac{D}{2}} (D+2+\rho^{-2})}{(1+\rho^{-2})^2-1}.
\label{FormFactorSimp} 
\end{equation} 
In the limit of $\lambda >> \xi$, or equivalently $\rho \rightarrow 0$, $F_D \rightarrow 1$ and the average total energy at equilibrium becomes 
\begin{equation}  
\langle E \rangle =\frac{D\left(N_A +N_B \right)}{\beta} +\gamma N_A N_B.
\end{equation}

In this limit we recover the behavior of the Caldeira-Leggett model. In particular, for the specific setting in which the model
is usually applied, with one of the two species playing the role of a large reservoir (for instance if $N_A>>N_B$), the
total energy becomes extensive, while being dependent on $N^2$ in the case of a balanced mixture ($N_A=N_B=N)$. 
The latter result is consistent with the idea that long range interactions are not extensive, as there will be
$N^2$ distinct interparticle interaction energy terms.

We now consider the situation where the thermal lengths are much larger than the interaction range, that
is $\lambda << \xi$ ($\rho >> 1$). In this regime the form factor $F_D$ depends on temperature, as seen in
Fig. \ref{Fig6}, and may be approximated as $F_D \simeq (1+D/2)(2 \rho^2)^{-D/2}$, with the corresponding 
expression for the average total energy

\begin{equation}  
\langle E \rangle \simeq \frac{D\left(N_A+N_B \right)}{\beta}
+\gamma \left(1+\frac{D}{2}\right) \left(\frac{\beta}{2\beta_\lambda}\right)^{\frac{D}{2}} N_A N_B.
\label{EquationTotalEnergy}  
\end{equation}

In the 1 D case and balanced mixtures ($N_A=N_B=N$), the average total energy 

\begin{equation}
\langle E \rangle \simeq \frac{2 N}{\beta} + \gamma \frac{3N^2}{2\sqrt{2}} \frac{\lambda}{\xi}
= \frac{2 N}{\beta} + \gamma \frac{3N^2}{2\sqrt{2}} \left(\frac{\beta}{\beta_\lambda}\right)^{1/2}. 
\end{equation}

The two terms constituting the average total energy depend linearly and quadratically on the number of
particles, respectively. We now impose a ``generalized extensivity'' property such that
$\langle E \rangle$ scales as $N^\alpha$, where the exponent $\alpha$ should lie
between the genuine extensive case of $\alpha=1$ achieved in the noninteracting case
and $\alpha=2$ reached in the strong coupling limit of $\gamma \rightarrow \pm\infty$.
This homogeneity in the two contributions to the total energy is achieved if the
inverse temperature itself depends on $N$ with a power-law exponent, more precisely if $\beta \propto N^{-\tau}$.
Then the two terms on the right-hand side will depend on $N^{1+\tau}$ and $N^{2-\tau/2}$, respectively.
The request for homogeneity is fulfilled if $\tau=2/3$. The average total interaction energy then
will scale as $\langle E_{\mathrm{int}} \rangle \sim N^{5/3}$, {\it i.e.} $\alpha=5/3$.
The evaluation of the scaling exponent is readily extended to $D$-dimensions, based on the second term of the righthand side
of Eq.~(\ref{EquationTotalEnergy}) and, using the same reasoning as above, we find the scaling exponent to depend on dimensionality as
\begin{equation}
\alpha=\frac{D+4}{D+2}.
\end{equation}
This means $\alpha=5/3, 3/2, 7/5$ in 1D, 2D and 3D, respectively. It is worth noticing that the extensive case is
obtained in the limit of infinite dimensions, and that quadratic scaling corresponds to a zero-dimensional system.
In order to compare these expectations with numerical simulations, one should add, on top of the request for a Maxwell-Boltzmann
distribution (which implies a sort of weak-coupling limit, with small values of $\gamma$) also the ergodic theorem in which the
ensemble averages evaluated above are matched by time-averaged quantities.
This is a requirement for thermal equilibration, as discussed in~\cite{JauOnoSun2}.

The scaling argument provided above may be considered as a necessary, but not sufficient, condition for
 the stability of the system. More insights on the stability with respect to the sign and the magnitude
of the interaction strength $\gamma$ may be arrived at by thermodynamic considerations.
In a stable thermodynamic system the entropy is a concave function of energy~\cite{Naudts}, 
which is always satisfied if the heat capacity is positive-valued.
In our case the heat capacity for short range interactions,  where $K_B$ is the Boltzmann constant, is
\begin{eqnarray}  
& & C=\frac{d\langle E\rangle }{dT} = D K_B \times \nonumber \\
& & \left[N_A+N_B-\frac{\beta \gamma}{2} \left(1+\frac{D}{2}\right){\left(\frac{\beta}{2\beta_\lambda}\right)}^{D/2}N_A N_B \right].
\label{HeatCapacity}
\end{eqnarray}
A change in the sign of the curvature in the entropy 
is indicative of a drastic change in the dynamical behavior, a sort of phase transition.
When the interaction is attractive ($\gamma <0$)  the heat capacity is always positive.
However for a repulsive interaction ($\gamma >0$) there exists a critical inverse temperature above which the system
is unstable. This threshold is given by
\begin{equation}
  \beta_{\mathrm{crit}}= 2 \left[\beta_\lambda \gamma \left(1+\frac{D}{2}\right)
\frac{N_A N_B}{N_A+N_B} \right]^{\frac{-2}{D+2}} \beta_\lambda
\label{Critical} 
\end{equation} 
The existence of a threshold can be simply understood by inspecting the motion of two generic interacting 
particles in the 1D case. Below the critical inverse temperature, both particles are free to explore the entire trap
while, at lower temperatures, each particle is confined on one side of the trap.
This can be thought of in terms of a phase separation which diminishes the interaction energy contribution, and
the overall scaling with the number of particles, in analogy to the discussion appeared in Sect.III of~\cite{OnoSun} in terms of stability analysis. At the critical inverse temperature and for an unbalanced mixture, the
total interaction energy is given by
\begin{equation}  
\langle E_{\mathrm{int}} \rangle \sim \gamma ^{2/(D+2)} \left(N_A N_B \right)^{2/(D+2)} \left(N_A +N_B \right)^{D/(D+2)},
\label{EintScal}
\end{equation} 
which obviously becomes extensive in one of the two systems when the other is composed of just one particle. For balanced
mixtures, the scaling confirms what was shown earlier, namely $\langle E_{\mathrm{int}} \rangle \sim N^{\frac{D+4}{D+2}}$. We note
that for our parameters (which involve balanced mixtures), the critical values fall within the inverse temperatures we consider
for the two species. Further, we stress that the estimate is valid only in the thermodynamic (large particle number) limit and that
we expect deviations given the small number of particles we consider.

In relation to an earlier comment on Fig.~\ref{Fig4} about the ratio of the interaction energy divided by $\gamma$ in the
limit $\gamma \rightarrow 0$, we need to extend the scaling relation (in $N$) to include the effects of $\gamma$ and $\lambda$.
The equilibrium temperature reached can be reasonably expected to depend on the strength $\gamma$ and the
range $\lambda$ of the interaction. In keeping with the earlier analysis, we consider
$T \propto N^{\alpha} \gamma^{\delta} \lambda^{\eta}$ and using analogous dimensional arguments, 
it can be shown that $\alpha=\delta= 2/(D+2)$ while $\eta=2D/(D+2)$.
Thus, for fixed $N$ and $\lambda$, the interaction energy scales as $\gamma^{1-\beta D/2}= \gamma^{2/(D+2)}$ (consistent
with Eq.~\ref{EintScal} derived from independent considerations) or $\gamma^{2/3}$ in one dimension.
This clearly indicates that the interaction energy goes to $0$ as $\gamma$ approaches $0$ from either
direction while the ratio shown in Fig.~\ref{Fig4} is ill-defined as $\gamma \rightarrow 0$.

\section{Concluding Remarks}

We have elaborated on scaling behavior, first reported in~\cite{JauOnoSun1}, seen 
in the interaction energy, of a binary mixture with short-range interactions, with respect to system size, at
the onset of thermalization. 
Contrasting extensive numerical simulations with analytic constructs we find that the scaling exponent
that coincides with the one seen in turbulent mixing occurs only for small positive values of the interaction coupling strength.
This is the regime where the interspecies interaction can be considered as a small perturbation
with respect to the external harmonic potential experienced by both species.
The scaling behaviors in other parameter regimes are more readily anticipated using simple analytic arguments.
It should be noted that scaling is also expected to break down when using nonlinear trapping potentials, where thermalization itself is also more involved, as discussed in~\cite{JauOnoSun2}.

Our results may have relevance in a variety of many-body physics contexts, including ultracold atomic physics
where the turbulent cascade of energy has been recently studied  both theoretically~\cite{Bradley} and
experimentally~\cite{Navon}, requiring extension of our model to the quantum realm. 
Although plasmas contain both intraspecies and interspecies interactions, the interplay among 
strong coupling, scaling behavior, and turbulence discussed here may be of interest
in the context of extremely exothermic systems such as magnetically confined fusion plasmas.
Features of plasmas can be isolated and simulated, in the spirit of the numerical studies for evaluating nuclear reaction
rates reported in~\cite{Dubin}. In particular, the relationship between Kolmogorov scaling and effective dimensionality
of confinement is crucial in magnetized fluids~\cite{Kraichnan,Antar,Mason,Perez}, and we plan to analyze scaling features
in the general case of anisotropic harmonic trapping.
Our model is also relevant to study efficient and fast heating, for instance, transferring to the plasma physics
context techniques developed for fast cooling in ultracold atomic physics~\cite{Chen,ChoOnoSun,Schaff,OnoRev,Diao}.
Additionally, the Caldeira-Leggett model has been shown to share similarities with the linearized Vlasov-Poisson equation,
including the presence of an analog of Landau damping~\cite{Morrison}. Our generalization of the Caldeira-Leggett model
to a nonperturbative setting should allow for the exploration of this analogy in a fully nonlinear regime, which is presumably
more appropriate for the description of plasma dynamics.


\begin{thebibliography}{99}

\bibitem{Magalinskii} V. B. Magalinskii, Sov. Phys. JETP \textbf{9}, 1381 (1959).

\bibitem{Ullersma} P. Ullersma, Physica \textbf{32}, 27 (1966); \textbf{32}, 56 (1966); \textbf{32}, 74 (1966); \textbf{32}, 90 (1966).

\bibitem{Caldeira1} A. O. Caldeira and A. J. Leggett, Phys. Rev. Lett. \textbf{46}, 211 (1981).

\bibitem{Caldeira2} A. O. Caldeira and A. J. Leggett, Ann. Phys. \textbf{149}, 374 (1983).

\bibitem{Caldeira3} A. O. Caldeira and A. J. Leggett, Phys. Rev. A \textbf{31}, 1059 (1985).

\bibitem{OnoSun} R. Onofrio and B. Sundaram, Phys. Rev. A \textbf{92}, 033422 (2015).

\bibitem{JauOnoSun1} F. Jauffred, R. Onofrio and B. Sundaram, J. Phys. B: At. Mol. Opt. Phys. \textbf{50}, 135005 (2017).

\bibitem{JauOnoSun2} F. Jauffred, R. Onofrio and B. Sundaram, Phys. Lett. A \textbf{381}, 2783 (2017).

\bibitem{Smith} S. T. Smith and R. Onofrio, Eur. Phys. J. B \textbf{61}, 271 (2008).

\bibitem{Khilchin} A.I. Khinchin, {\sl Mathematical Foundations of Statistical Mechanics} (Dover Publications Inc., New York, 1949).

\bibitem{Montroll} E. W. Montroll and K. E. Shuler, J. Chem. Phys. \textbf{26}, 454 (1957).

\bibitem{AndersenK} K. Andersen and K. E. Shuler, J. Chem. Phys. \textbf{40}, 633 (1964).

\bibitem{AndersenH} H. C. Andersen, I. Oppenheim, K. E. Shuler and G. H. Weiss, J. Math. Phys. \textbf{5}, 522 (1964).

\bibitem{Naudts} J. Naudts, J. Phys.: Conf. Ser. \textbf{201}, 012003 (2010).

\bibitem{Bradley} A. S. Bradley and B. P. Anderson, Phys. Rev. X \textbf{2}, 041001 (2012).

\bibitem{Navon} N. Navon, A. L. Gaunt, R. P. Smith and Z. Hadzibabic, Nature \textbf{539}, 72 (2016).  
  
\bibitem{Dubin} D. H. E. Dubin, Phys. Plasmas \textbf{15}, 055705 (2008).
  
\bibitem{Kraichnan} R. H. Kraichnan, Phys. Fluids \textbf{8}, 1385 (1965).
  
\bibitem{Antar} G. Y. Antar, Phys. Rev. Lett. \textbf{91}, 055002 (2003). 

\bibitem{Mason} J. Mason, F. Cattaneo and S. Boldyrev, Phys. Rev. Lett. \textbf{97}, 255002 (2006).

\bibitem{Perez} J. C. Perez, J. Mason, S. Boldyrev and F. Cattaneo, Phys. Rev. X \textbf{2}, 041005 (2012).  

\bibitem{Chen} X. Chen, A. Ruschhaupt, S. Schmidt, A. del Campo, D. Gu\'ery-Odelin and J. G. Muga, 
Phys. Rev. Lett. \textbf{104}, 063002 (2010).

\bibitem{ChoOnoSun}  S. Choi, R. Onofrio and B. Sundaram, Phys. Rev. A \textbf{84}, 051601(R) (2011).

\bibitem{Schaff} J.-F. Schaff, P. Capuzzi, G. Labeyrie and P. Vignolo, New. Journ. Phys. \textbf{13}, 113017 (2011).

\bibitem{OnoRev} R. Onofrio, Phys. Uspekhi \textbf{59}, 1129 (2016). 
  
\bibitem{Diao} P. Diao, S. Deng, F. Li, S. Yu, A. Chenu, A. del Campo and H. Wu, New. J. Phys. \textbf{20}, 105004 (2018).

\bibitem{Morrison} G. I. Hagstrom and P. J. Morrison, Physica D \textbf{240}, 1652 (2011).

\end{thebibliography}
\end{document}